\makeatletter
\declare@file@substitution{revtex4-1.cls}{revtex4-2.cls}
\makeatother
\documentclass[twocolumn, times, tighten,twocolappendix, a4paper]{aastex63}
\usepackage{amsmath,amssymb}
\usepackage[T1]{fontenc}
\usepackage{graphicx,multirow,hyperref,url,color,xspace}
\usepackage{lineno}

\usepackage{rotating}
\usepackage{subfigure}

\newcommand{\msun}{{M}_{\sun}}

\newcommand{\nustar}{{NuSTAR}\xspace}
\newcommand{\nicer}{NICER\xspace}

\newcommand{\source}{{GX 339--4}\xspace}

\newbox\grsign \setbox\grsign=\hbox{$>$} \newdimen\grdimen \grdimen=\ht\grsign
\newbox\simpropbox
\setbox\simpropbox=\hbox{\raise.5ex\hbox{$\propto$}\llap
 {\lower.5ex\hbox{$\sim$}}}\ht2=\grdimen\dp2=0pt

\begin{document}

\title{Is the Spin of the Black Hole in GX 339--4 Negative?}

\author[0000-0002-0333-2452]{Andrzej A. Zdziarski}
\affiliation{Nicolaus Copernicus Astronomical Center, Polish Academy of Sciences, Bartycka 18, PL-00-716 Warszawa, Poland; \href{mailto:aaz@camk.edu.pl}{aaz@camk.edu.pl}}

\author[0000-0002-6051-6928]{Srimanta Banerjee}
\affiliation{Nicolaus Copernicus Astronomical Center, Polish Academy of Sciences, Bartycka 18, PL-00-716 Warszawa, Poland;
\href{mailto:aaz@camk.edu.pl}{aaz@camk.edu.pl}}

\author[0000-0001-7606-5925]{Micha{\l} Szanecki}
\affiliation{Faculty of Physics and Applied Informatics, {\L}{\'o}d{\'z} University, Pomorska 149/153, PL-90-236 {\L}{\'o}d{\'z}, Poland}

\author[0000-0002-7609-2779]{Ranjeev Misra}
\affiliation{Inter-University Center for Astronomy and Astrophysics, Pune 411007, India}

\author[0000-0003-1589-2075]{Gulab Dewangan}
\affiliation{Inter-University Center for Astronomy and Astrophysics, Pune 411007, India}

\begin{abstract}
We have studied the accreting black hole binary GX 339--4 using two highly accurate broad-band X-ray data sets in very soft spectral states from simultaneous NICER and NuSTAR observations. Joint fitting of both data sets with relativistic models of the disk, its Comptonization and reflection allows us to relatively accurately determine the black-hole mass and spin, and the distance and inclination. However, we find the measured values strongly depend on the used disk model. With widely used Kerr disk models treating departures from local blackbody spectra using color corrections, we find relatively low black-hole masses and strongly negative spins (i.e., retrograde accretion). Then, models employing radiative transfer calculations of the disk atmosphere predict moderately positive spins and high masses. When adding a warm corona above the disk (as proposed before for both AGNs and accreting binaries), we find the spin is weakly constrained, but consistent with zero. In all cases, the fitted inclination is low, $\approx$30--$34\degr$. For the spin axis aligned with the binary axis, the mass function for this source implies large values of the mass, consistent only with those obtained with either disk-atmosphere models or the presence of a warm corona. We also test different disk models for an assumed set of mass, distance and inclination. We find that different models yield values of the spin parameter differing up to $\sim$0.3. Our results confirm previously found strong model dependencies of the measured black-hole spin, now by comparing different disk models and for a low-mass X-ray binary.
\end{abstract}

\section{Introduction} \label{intro}

\source is a well-studied low-mass X-ray binary (LMXB), whose accretor is most likely a black hole (BH; \citealt{Heida17}). Its BH mass, $M_1$, distance, $D$, and inclination, $i$, are poorly constrained \citep{Heida17, Zdziarski19a}. The spin of the black hole in this source was estimated in the hard spectral state (in which the accretion disk is weak and the X-ray spectrum is dominated by a hard component, apparently due to thermal Comptonization) using X-ray reflection spectroscopy \citep{Bambi21}, which relies on relativistic broadening of spectral features from reflection (mostly the fluorescent Fe K$\alpha$ line). Studies with that method \citep{Reis08, Miller08, Ludlam15, Garcia15} found the BH to be close to maximally spinning. However, the validity of those results requires that the accretion disk in the hard state extends very close to the innermost stable circular orbit (ISCO), which remains highly uncertain in the hard state in general (e.g., \citealt{DGK07}), and in \source in particular \citep{DeMarco15, Basak16, Dzielak19, Mahmoud19, ZDM20}. When the inner disk radius is far from the ISCO, the space-time metric is insensitive to the spin, and it cannot be determined.

A more reliable method appears to be continuum fitting \citep{McClintock14} in the soft spectral state, whose X-ray spectra are dominated by an optically-thick accretion disk. The method relies on the dependence of the ISCO radius on the spin \citep{Bardeen72}, and it uses relativistic models of accretion disks to fit observed X-ray spectra. However, this method is usually reliable only when the mass, distance and the inclination are known, which is not the case for \source. Some studies of the soft state were performed \citep{Kolehmainen10, Kolehmainen11}, but they were able only to constrain the spin parameter to $a_*\lesssim 0.9$. Still, spectral fitting of high-quality broad-band data in this state can potentially break the degeneracies between the mass, distance and inclination inherent to relativistic disk models. A study with such a goal was done by \citet{Parker16} for the very high state of \source. That state consists of a strong blackbody disk spectrum joining smoothly on a strong power law, most likely from Compton scattering. In our view, a spectral decomposition of such relatively complex total spectrum is prone to large systematic uncertainties\footnote{We note that the model of \citet{Parker16} included a strong component from thermal Comptonization by a plasma with a Thomson optical depth of $\tau_{\rm T}\lesssim 0.01$ scattering on disk blackbody photons.  However, such a plasma scatters only a $\tau_{\rm T}$ fraction of the incident flux of seed photons, and it would thus produce a spectrum with a normalization two orders of magnitude below that found in their spectral fit.}.

Here, we pursue the same goal by studying two data sets of simultaneous NICER and NuSTAR observations of this source in very soft states, i.e., strongly dominated by disk blackbodies and with very weak high-energy tails. Both data sets are of outstanding statistical quality as well as their spectral calibrations agree very well with each other in the overlapping energy ranges. We use several different models of optically-thick relativistic disks accreting onto BHs. The models differ in their treatment of modifications of the disk spectra with respect to the LTE. Some treat them using a color correction, i.e., the ratio of the color temperature to the effective one. Then, some use calculations of the vertical radiative transfer of the disk emission. Furthermore, some assume the disk to be geometrically thin, while some take into account the finite scale height (so-called slim disks), increasing when the luminosity becomes comparable to the Eddington value.

We then compare the fitted parameters between different models and between our two data sets. We find the results strongly differ between the used disk models for a given data sets, and, in some cases, between the two data set for a given model. We attempt to reconcile those differences and to find the most likely set of the source parameters.

\section{The data and variability}
\label{data}

We consider two NICER/NuSTAR data sets in X-rays from 2020 and 2021. The log of the observations is given in Table \ref{log}. Each spectrum consists of an apparent disk blackbody and a weak high-energy tail with the photon index of $\Gamma\sim 2$. The data are of very high statistical quality, and are strongly dominated by the disk blackbody components. The spectra from NICER, NuSTAR A and NuSTAR B agree very well with each other in the overlapping ranges of the energy. We normalize the spectra to those of NuSTAR A. The differences in the absolute normalization between NuSTAR A and B was about 3\% (in agreement with \citealt{Madsen22}), and those between NICER and NuSTAR A, about 8 and $<1$\% for the 2020 and 2021 data, respectively. The difference between those factors for the two data sets appears to be mostly due to the observations by NICER and NuSTAR not being completely simultaneous, see Table \ref{log}.

The \nicer spectra are used in the 0.5--6.8 and 0.5--10 keV for the 2020 and 2021 data, respectively. The former data are strongly background-dominated above 7 keV. The second good-time-interval of the NICER 2020 light curve was excluded due to the occurrence of flares caused by a non-X-ray event. During the reduction pipeline of NICER, the script {\tt nicerl3-spect} added 1.5\% systematic error to the detector channels in the 0.5--9.0\,keV range, and more (up to 2.5\%) above 9 keV\footnote{\url{https://heasarc.gsfc.nasa.gov/lheasoft/ftools/headas/nicerl3-spect.html}}. The \nustar data are used in the $\approx$3--50 keV range, with the data at $>50$ keV being strongly background-dominated and thus not usable. They were extracted using the option for bright sources\footnote{{\tt statusexpr=\\"(STATUS==b0000xxx00xxxx000)\&\&(SHIELD==0)"}.}. The spectral data have been optimally binned \citep{Kaastra16} with an additional requirement of at least 20 counts per bin. 

The variability at the disk-dominated energies was very weak. The fractional rms for the 2021 observation at 0.01--1 Hz frequency range and in the 0.5--3 keV (from NICER) and 3--10 keV (from NuSTAR) energy ranges are $\approx 0.39\pm 0.07\%$ and $1.05\pm 0.45\%$, respectively. The 3--10 keV rms larger than that of 0.5--3 keV is likely due to a contribution from a variable Comptonization component, see the spectrum, which is dominated by the tail at $>$8 keV (Section \ref{results} below). Thus, the disk was very stable with a more variable high-energy tail. Then, the rms for the 2020 observation was so low that it could not be reliably estimated. Thus, the disk was very stable in that case as well. The tails above 10 keV were so weak that their rms could not be reliably estimated as well. 

\begin{table*}
\centering
\caption{The log of the observations. \label{log}}
  \begin{tabular}{cccccccc}
    \hline
NICER & Start Time & Exposure (s) & NuSTAR & Start Time  & Exposure (s) & Exposure (s)\\
Obs. ID    &  End Time  &  & Obs. ID & End Time & FPM A & FPM B\\
 \hline
2635010101 & 2020-02-20 03:39:00 & 2384 & 80502325002 & 2020-02-20 03:06:09 & 17820 & 17829\\
  & 2020-02-20 20:48:20 &  &  & 2020-02-20 14:06:09 & \\
4133010131 & 2021-04-24 01:55:40 & 3441 & 90702303013 & 2021-04-23 19:41:09 & 20046 & 20609\\
  & 2021-04-24 13:03:00 &  &  & 2021-04-24 09:31:09&\\
\hline
 \end{tabular}
 \end{table*}

\section{Models}
\label{models}

In our main approach, we follow the self-consistent modelling of the soft state of Cyg X-1 of \citet{Zdziarski24b}, using a disk model, its Comptonization, and the reflection spectrum of the Comptonized emission incident on the disk modeled by the convolution model {\tt xilconv} \citep{Kolehmainen11, Garcia13, MZ95}, which is then relativistically broadened using {\tt relconv} \citep{Dauser10}. For the latter, we constrain the irradiation index (defined by $d F_E\propto r^{-q}$) to $q\leq 6$, in order to avoid an extreme concentration of the reflection from the immediate vicinity of the ISCO. We link the inclination and the spin between the relativistic disk and the reflection components. For the Comptonization, we use now a convolution version of the {\tt compps} model of \citet{PS96}, {\tt comppsc}\footnote{\url{https://github.com/mitsza/compps_conv}}, as used in the modelling of LMC X-1 in \citet{Zdziarski24a}. In our fits, we use the Compton parameter, $y\equiv 4 \tau_{\rm T} kT_{\rm e}/m_{\rm e}c^2$ (where $kT_{\rm e}$ is the electron temperature in energy units), instead of the Thomson optical depth, $\tau_{\rm T}$. The former determines the slope of power-law parts of the Comptonization spectra. We have tested both a slab and spherical geometries, the latter using the option {\tt geom=0}, which is based on a method using escape probabilities. We have found only small differences in the main fitted parameters between the two approaches, which we quantify in Section \ref{slimbh} below. Thus, we opt for the latter, whose calculations are much faster than in the former case. 

While the considered data sets have relatively weak high-energy tails, their fitting did require the presence of a nonthermal, power-law, tail in the coronal electron distribution. The tail has the form $(\gamma\beta)^{-p}$ (where $\beta$ is the dimensionless electron velocity) between $\gamma_{\rm min}$, at which Lorentz factor the Maxwellian and the power law join, up to $\gamma_{\rm max}$. In order to check possible effects of changing the treatment of Comptonization, we have also used the convolution thermal Comptonization model {\tt thcomp} \citep{Z20_thcomp}. We have generally found similar results to those using {\tt comppsc}, though the fits to the high-energy tails were worse. Since {\tt thcomp} assumes $\tau_{\rm T}>1$ only and it does not allow for the presence of nonthermal electrons, we decided to only use {\tt comppsc} in our presented models. 

Among the considered disk models, we begin with {\tt diskbb} \citep{Mitsuda84}, which is a simple non-relativistic disk model. While it does not allow us to estimate the spin, it gives an approximate estimate of the disk inner radius. Next, we use a number of codes for relativistic disk modelling. The first is {\tt kerrbb} \citep{Li05}, which is a model of a geometrically thin disk \citep{NT73}. Atmospheric modifications of the blackbody emission are handled via a color correction, $f_{\rm col}$. Its typical values in standard X-ray disk models are 1.4--2.0 \citep{Davis19}. We allow $f_{\rm col}$ to be free and $\leq 2.0$ as advocated by \citet{Salvesen21}. We note that in the presence of large scale magnetic fields $f_{\rm col}$ could be $>2$ \citep{Begelman07}, but we neglect that case. The next one is {\tt kerrbb2} \citep{McClintock06}, which is a modification of {\tt kerrbb} with the color corrections fitted to the results of the disk-atmosphere calculations by \citet{Davis05} and \citet{Davis06}. Then, the model {\tt bhspec} uses directly those calculations instead of a color correction. All of those relativistic disk models assume the inner radius to be at the ISCO. We note that {\tt kerrbb} and {\tt kerrbb2} use the mass accretion rate, $\dot M$, as a parameter, while {\tt bhspec} (and {\tt slimbh}, see below) use the Eddington ratio (assuming pure H). The latter is related to the former by
\begin{equation}
\frac{L}{L_{\rm E}}=\frac{\eta(a_*)\dot M c \sigma_{\rm T}}{4\pi G M_1 m_{\rm p}},
\label{eddratio}
\end{equation}
where $L_{\rm E}$ is the Eddington luminosity (which we define here for pure H), $\eta(a_*)$ is the accretion efficiency \citep{Bardeen72}, $M_1$ is the BH mass, $\sigma_{\rm T}$ is the Thomson cross section and $m_{\rm p}$ is the proton mass. 

\begin{figure*}[t!]
\centerline{\includegraphics[width=7.5cm]{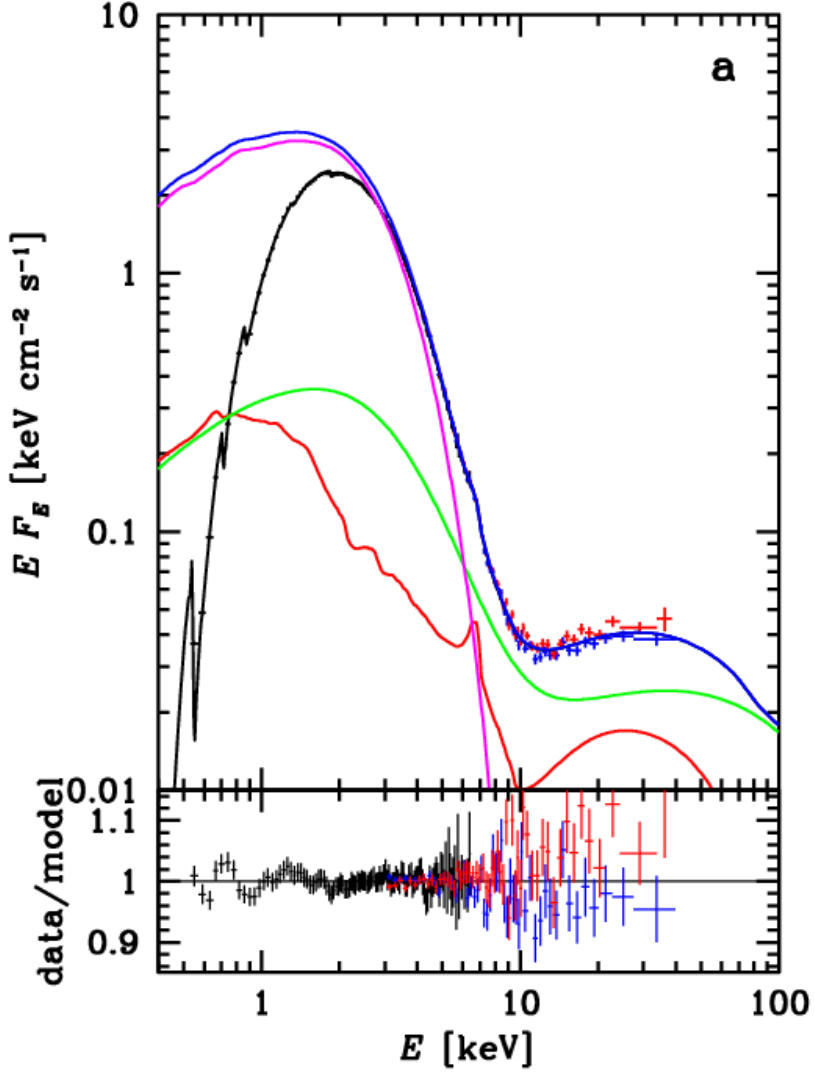}
\includegraphics[width=7.5cm]{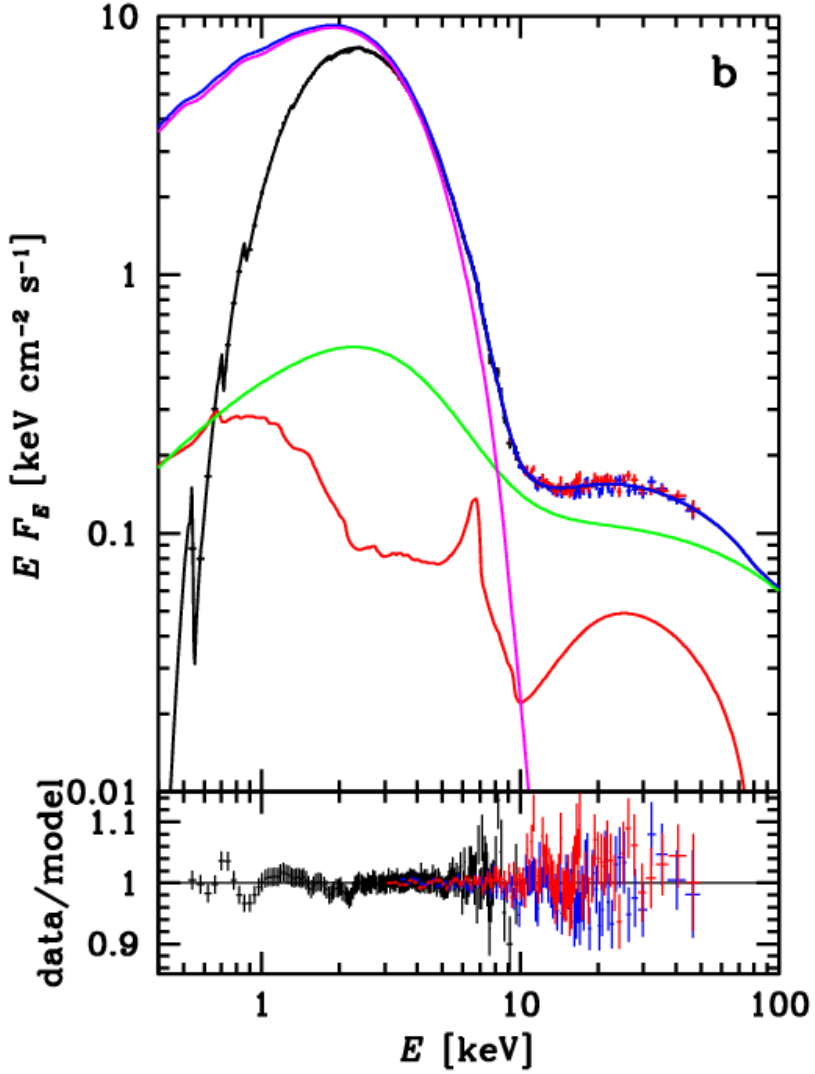}}
\caption{The \nicer (black) and \nustar (blue and red) unfolded spectra (top panels) and data-to-model ratios (bottom panels) for the joint fit with model 3 (see Table \ref{fits}) to the data from (a) 2020 and (b) 2021. The spectra are normalized to \nustar A. The model is of Equation (\ref{xilconv}) with {\it disk} = {\tt slimbh} and the atmospheric spectra. The total model spectra and the unabsorbed one are shown by the solid black and blue curves, respectively. The unabsorbed disk emission, scattering alone and reflection are shown by the magenta, green, and red curves, respectively. 
}\label{f_models}
\end{figure*}

Another refinement consists of taking into account the finite disk thickness, important when the disk luminosity becomes $\gtrsim 0.1 L_{\rm E}$, which is the case for the 2021 data. The disk then becomes 'slim' \citep{Abramowicz88}. In order to account for it, we use the model based on calculations of \citet{Sadowski09, Sadowski11a} and \citet{Sadowski11b}, {\tt slimbh} (described in \citealt{Straub11}). Modifications to the blackbodies are treated either by a color correction, or by atmospheric calculations, based on the results of \citet{Davis05} and \citet{Davis06}. Note that the atmospheric radiative transfer calculations in {\tt bhspec} and {\tt slimbh} assume disks with magnetic fields generated only by the standard MRI, i.e., not magnetically dominated. 

The above models are of the form
\begin{align}
&{\tt constant*tbfeo*comppsc}\{{\it disk}+\nonumber\\
&{\tt relconv[xilconv(comppsc}({\it disk}))]\},
\label{xilconv}
\end{align}
where {\it disk} stands for either {\tt diskbb}, {\tt kerrbb}, {\tt kerrbb2}, {\tt bhspec} or {\tt slimbh}. Then {\tt constant} accounts for differences in the flux normalization of the different detectors. Here, the first term in the external parentheses gives the disk emission, and the second term gives the reflection of the scattered radiation. Then both of them undergo scattering in the corona, as given by first appearance of {\tt comppsc}. The second term uses the option giving the scattered spectrum only. Thus, that term consists of reflection of the scattered disk emission that is then again Compton scattered during crossing the corona. 

For the interstellar absorption, modelled by {\tt tbfeo} \citep{Wilms00}, we assume the abundances of \citet{AG89}, which consistently gave better fits at low energies than those of \citet{Wilms00}. However, we allow for the abundances of O and Fe to differ from solar, which we found was necessary for fitting the NICER data. We quantify it in Section \ref{slimbh}. We set the photon energy grid, which is necessary for convolution models, by the {\sc xspec} command {\tt energies 0.01 100 1000 log}, except that for {\tt bhspec}, which is tabulated at $E\geq 0.1$ keV only, we use {\tt energies 0.1 100 1000 log}.

We also tested changing the treatment of the reflection and its relativistic broadening. For that, we used the reflection model in which the incident photons form a thermal Comptonization spectrum \citep{Garcia18}, called {\tt xillverCp}. Its version including the relativistic broadening is {\tt relxillCp}. We use it for the reflected spectrum only. Models using {\tt relxillCp} are then of the form 
\begin{align}
&{\tt constant*tbfeo*comppsc({\it disk}+}\nonumber \\
&{\tt mbkpno*relxillCp)},\label{relxill}
\end{align}
where {\tt mbkpno} \citep{Svoboda24} imposes a low-energy break at an energy, $E_{\rm b}$, on the reflection spectrum from {\tt relxillCp}. Specifically, it is equal to 1 at $E\geq E_{\rm b}$, and $(E/E_{\rm b})^r$ at $E>E_{\rm b}$. Here, we assume $r=1$. Such a factor is needed because the incident spectrum of that model assumes blackbody seed photons at the low temperature of 50 eV. However, {\tt comppsc} has the Compton $y$ as the parameter specifying the slope of the scattered spectrum while {\tt relxillCp} uses the photon index $\Gamma$, as well as the normalization of it can be arbitrary, and thus not related to the Comptonizing flux incident on the disk. Testing it, we found that while for some cases the above model reproduces well the results of the model of Equation (\ref{xilconv}), it gives different, and strongly unphysical, results in some other cases. Furthermore, we have found that the Comptonization incident spectra in the soft state can have the shapes very different from a power with a high-energy cutoff, see Section \ref{slimbh} below. This is especially the case for the 2020 spectrum. Thus, we present mostly results using Equation (\ref{xilconv}). 

Finally, we consider a possibility that the accretion disk in the soft state of BH X-ray binaries (XRBs) can be covered by a top warm scattering layer (a warm corona), as in a popular model for accretion disks in AGNs (e.g., \citealt{Petrucci20, Ballantyne24}), where the warm coronae explain the AGN soft X-ray excesses. Such warm coronae were also found necessary to explain the optical-to-X-ray spectra of AGNs at medium and high Eddington ratios \citep{Hagen24, Kang24}. In the studies of the soft-state spectra of Cyg X-1, LMC X-1 and M33 X-7 \citep{Belczynski24, Zdziarski24a, Zdziarski24b}, the warm corona was modeled by optically-thick thermal Comptonization using {\tt thcomp}. Those models yielded low spins for those objects, consistent with $a_*\sim 0.1$. Here, we consider our model of Equation (\ref{xilconv}) modified by adding a warm layer fully covering the disk for its most advanced model considered here, {\tt slimbh},
\begin{align}
&{\tt constant*tbfeo*comppsc\{thcomp(slimbh)+}\nonumber\\
&{\tt relconv[xilconv(comppsc(thcomp(slimbh))))]\} }.
\label{warm}
\end{align}

\citet{Zdziarski19a} estimated the distance as $8\,{\rm kpc}\leq D\leq 12\,$kpc, which we assume hereafter. The shape of the track of \source on the X-ray hardness-count rate diagram implies $i\lesssim 60\degr$ \citep{Munoz13}. We assume $M_1\geq 4\msun$. The reflection fraction, ${\cal R}$, in the case of Equation (\ref{xilconv}) is limited to 2.0. Keeping it free within that limit accounts for model uncertainties, e.g., the likely difference between the outgoing flux and that incident on the disk. We use {\sc{xspec}} \citep{Arnaud96} for spectral fitting, and estimate the uncertainties calculated for 90\% confidence ($\Delta\chi^2 \approx 2.71$; \citealt{Lampton76}). 

\section{Results}
\label{results}

\subsection{Models with {\tt diskbb}}
\label{diskbb}

We first consider the simplest disk model, {\tt diskbb}, with the Comptonization continuum and reflection modelled as in Equation (\ref{xilconv}) with {\it disk} = {\tt diskbb}. We find the maximum disk temperatures for the 2020 and 2021 data are well constrained to $0.62$ and $0.82$ keV, respectively. The normalization and inclination are similar for both data sets, $N_{\tt dbb}\approx 3080_{-150}^{+100}$ and $2920_{-70}^{+30}$, $i\approx 33^{+5}_{-3}\degr$, $34^{+1}_{-2}\degr$, and the fits are very good, $\chi_\nu^2\approx 323/334$ and 447/469, respectively. We see that both $N_{\rm diskbb}$ and $i$ are compatible with being the same for both data. 

We then fit the two data sets jointly, assuming the ISM absorption, $i$, $a_*$, $N_{\tt dbb}$, and the Fe abundance, $Z_{\rm Fe}$, to be identical for both data sets. We obtain a very good fit, $\chi_\nu^2\approx 772/811$, see Table \ref{fits} for the parameters, where it is denoted as Model 1. We note that this $\chi^2$ is larger than the sum of the $\chi^2$ for the individual fits by 2 only, which underlines the full consistency of the parameters fitted individually to the two spectra. Interestingly, the spin in this model, which is determined solely by the reflection, is strongly negative, $a_*=-1^{+0.45}$. We have tested a variant of this model using {\tt relxillCp}, Equation (\ref{relxill}). We have found that it yields a much worse fit, with $\chi_\nu^2=857/806$. 

\subsection{Models with {\tt kerrbb} and {\tt kerrbb2}}
\label{kerrbb}

We then consider models with {\tt kerrbb}. The model using Equation (\ref{xilconv}) gave $\chi_\nu^2= 327/332$, 471/467, $a_*=-1.00^{+1.99}$ and $-0.90_{-0.10}^{+0.23}$ for the two data sets, respectively. When both data sets are fitted jointly with {\tt kerrbb}, $\chi_\nu^2= 833/807$, and we find the negative spin of $a_*=-0.53^{+0.34}_{-0.47}$, with a monotonic dependence of $\chi^2$ on the increasing $a_*$, with $\Delta\chi^2=+137$ at $a=0.998$. See Table \ref{fits} for its other parameters, where it is denoted as model 2. Negative spins were also obtained when Equation (\ref{relxill}) was used. 

Then, we considered {\tt kerrbb2}. The model using Equation (\ref{xilconv}) gave $a_*=-0.80_{-0.20}^{+0.92}$ and $-1.00^{+0.28}$ for the two data sets, respectively. In both cases, $i=32\degr$, $D\approx 11$ kpc, and the BH mass is very low, $M_1\approx 4\msun$. The joint fit yields $a_*=-1^{+0.05}$, $M_1= 4.1_{-0.1}^{+0.2}\msun$, $D= 10.8_{-0.5}^{+0.2}$ kpc, at $\chi_\nu^2= 820/810$, see Table \ref{review}, where we compare the obtained spins for all of the disk models. The similarity of these results to those with {\tt kerrbb} is surprising in the light of our results in Section \ref{slimbh} below, in which we find that models utilizing directly the atmospheric spectra give strongly different results from those of {\tt kerrbb2}, and with large positive spins. This is in spite of {\tt kerrbb2} using the color corrections fitted to the same atmospheric calculations. 

\renewcommand{\arraystretch}{0.9}
\begin{table*}
\caption{The results of joint spectral fitting}
   \centering\begin{tabular}{lccccc}
\hline
Group/Model & Parameter & 1. {\tt diskbb} & 2. {\tt kerrbb} & 3. {\tt slimbh} & 4. warm\\
\hline
Joint & $N_{\rm H}$ $[10^{21}$\,cm$^{-2}]$ & $5.6^{+0.1}_{-0.1}$    & $5.7^{+0.1}_{-0.1}$& $6.0^{+0.1}_{-0.1}$& $5.6^{+0.1}_{-0.1}$  \\
parameters & $Z_{\rm O,abs}$ & $0.67^{+0.02}_{-0.02}$  & $0.65^{+0.03}_{-0.03}$  & $0.67^{+0.02}_{-0.03}$   & $0.67^{+0.03}_{-0.03}$  \\
  & $Z_{\rm Fe,abs}$& $0.77^{+0.04}_{-0.04}$  & $0.80^{+0.04}_{-0.04}$  & $0.90^{+0.04}_{-0.04}$   & $0.75_{-0.05}^{+0.04}$ \\    
  & $a_*$& $-1.00^{+0.46}$  & $-0.53^{+0.34}_{-0.47}$ & $0.71^{+0.05}_{-0.04}$   & $0.00^{+0.79}$  \\
  & $M_1\,[\msun]$& --& $6.9_{-2.8}^{+3.2}$& $12.5_{-0.3}^{+1.0}$& $5_{-1}^{+9}$\\
  & $N_{\tt dbb}$ or $D$\,[kpc]  & $2974_{-38}^{+50}$    & $10.6^{+1.4}_{-1.1}$  & $11.5^{+0.5}_{-0.8}$    & $8.5_{-0.5}^{+2.9}$\\
  & $i\,[\degr]$  & $32^{+1}_{-1}$  & $33^{+2}_{-1}$ & $32^{+2}_{-2}$& $33^{+1}_{-2}$ \\ 
  & $Z_{\rm Fe,disk}$   & $2.0^{+0.3}_{-0.7}$    & $6.0_{-1.5}$  & $3.6^{+1.2}_{-0.7}$& $1.6_{-0.5}^{+0.4}$ \\  
\hline  
Disk  & $kT_{\rm max,0}$ [keV] or $L_0/L_{\rm E}$ & $0.623^{+0.001}_{-0.001}$ & $0.2_{-0.1}^{+0.3}$ & $0.09^{+0.01}_{-0.01}$ & $0.08^{+0.03}_{-0.03}$  \\
  & $kT_{\rm max,1}$ [keV] or $L_1/L_{\rm E}$ & $0.813^{+0.002}_{-0.002}$ &$0.6_{-0.3}^{+0.4}$ & $0.24^{+0.02}_{-0.02}$ & $0.15^{+0.05}_{-0.06}$  \\
  & $f_{\rm col,0}$ & --   & $1.95^{+0.01}_{-0.14}$& --& $1.6^{+0.4}_{-0.2}$ \\
  & $f_{\rm col,1}$ & --   & $2.0_{-0.1}$& --& $1.3^{+0.1}_{-0.1}$  \\
\hline 
Comptonization  & $kT_{\rm e,0}$\,[keV]   & $11^{+1}_{-1}$  & $27^{+9}_{-6}$   & $17^{+2}_{-2}$  & $23^{+1}_{-1}$  \\
  & $kT_{\rm e,1}$\,[keV]   & $38^{+3}_{-5}$ & $78^{+11}_{-6}$  & $=kT_{\rm e,0}$  & $=kT_{\rm e,0}$ \\
  & $y_0$    & $0.12^{+0.03}_{-0.01}$    & $0.3^{+0.2}_{-0.2}$    & $0.06^{+0.02}_{-0.01}$  & $0.07^{+0.04}_{-0.02}$  \\
  & $y_1$    & $0.018^{+0.003}_{-0.002}$ & $0.013^{+0.002}_{-0.002}$  & $0.05^{+0.04}_{-0.02}$  & $0.03^{+0.01}_{-0.01}$  \\
  & $p_0$    & $1.4^{+0.4}_{-0.6}$ & $2.3_{-1.2}^{+0.6}$  & $0.9_{-0.8}^{+1.0}$&$0.04^{+0.91}_{-0.04}$\\
  & $p_1$    & $2.0^{+1.1}_{-0.2}$ & $1.7_{-0.5}^{+1.4}$& $2.6_{-0.1}^{+0.7}$    & $2.2^{+0.4}_{-0.1}$\\
  & $\gamma_{\rm min,0}$& $1.24^{+0.05}_{-0.03}$    & $1.5^{+0.1}_{-0.2}$& $1.4^{+0.1}_{-0.1}$    &$1.5^{+0.2}_{-0.1}$ \\
  & $\gamma_{\rm min,1}$& $1.6^{+0.1}_{-0.1}$ & $2.0^{+0.2}_{-0.2}$   & $1.26^{+0.01}_{-0.03}$  &$1.37^{+0.02}_{-0.02}$ \\
  & $\gamma_{\rm max,0,1}$& $6.2^{+2.8}_{-0.2}$ & $\geq$6.0  & $6.0^{+2.4}$  &$6.1^{+2.0}_{-0.1}$ \\
  &$f_{\rm cov,0}$ & $0.11^{+0.01}_{-0.03}$    & $0.02^{+0.01}_{-0.01}$& $0.4^{+0.1}_{-0.1}$    &$0.2^{+0.1}_{-0.1}$   \\
  &$f_{\rm cov,1}$ & $0.5^{+0.1}_{-0.1}$ &  $0.7^{+0.1}_{-0.1}$  & $0.3^{+0.1}_{-0.1}$    &$0.4^{+0.1}_{-0.1}$   \\
  & $\tau_{\rm warm,0}$ & --   & --  &--   &$15^{+10}_{-10}$  \\
  & $\tau_{\rm warm,1}$ & --   & --  &--   &$25_{-6}^{+2}$  \\
  & $kT_{\rm warm,0}$ [keV] & --   & --  & --  &$0.57^{+0.02}_{-0.14}$   \\
  & $kT_{\rm warm,1}$ [keV] & --   & --  & --  &$0.73^{+0.01}_{-0.01}$   \\
\hline 
Reflection & ${\cal R}_0$ & $1.1^{+0.2}_{-0.1}$ & $0.6^{+0.2}_{-0.1}$ & $0.7^{+0.1}_{-0.1}$    & $1.5^{+0.4}_{-0.4}$   \\
  &${\cal R}_1$  & $2.0_{-0.1}$   & $1.2^{+0.3}_{-0.2}$ & $0.5^{+0.1}_{-0.1}$    & $2.0_{-0.1}$\\  
  &$\log_{10}\xi_0$    & $4.3^{+0.1}_{-0.2}$ & $3.7^{+0.2}_{-0.2}$ & $4.3^{+0.1}_{-0.2}$    & $4.1^{+0.1}_{-0.1}$   \\
  &$\log_{10}\xi_1$    & $3.3^{+0.1}_{-0.1}$ & $2.7^{+0.1}_{-0.1}$ & $3.7^{+0.1}_{-0.2}$    & $3.7^{+0.1}_{-0.1}$   \\
  & $q_0$    & $2.8^{+0.4}_{-0.3}$ & $6.0_{-1.5}$ & $2.2^{+0.2}_{-0.2}$    & $3.0^{+0.8}_{-0.5}$ \\
  & $q_1$    & $3.4^{+0.2}_{-0.2}$ & $4.5^{+1.1}_{-1.0}$   & $=q_0$ &$=q_0$ \\
\hline 
  & $\chi_\nu^2$  & 772/811 & 833/808 & 765/812& 751/806 \\
\hline
\end{tabular}
\tablecomments{Models 1, 2 and 3 follow Equation (\ref{xilconv}) and model 4 follows Equation (\ref{warm}). The subscripts of 0, 1 for the disk, Comptonization and reflection groups correspond to the 2020 and 2021 data sets, respectively. We have constrained some parameters as $Z_{\rm Fe,disk}\leq 6.0$, $\gamma_{\rm min}\geq 1.2$, $\gamma_{\rm max}\geq 6.0$ and $\gamma_{\rm max,1}= \gamma_{\rm max,0}$, $p\geq 0$, ${\cal R}\leq 2$. In models 3, 4, $a_*\geq 0$ is a constraint of {\tt slimbh}. Model 3 is preferred when considering the fit quality, agreement with the mass function (Section \ref{preferred}) and its relative simplicity. }
\label{fits}
\end{table*}

\renewcommand{\arraystretch}{1.0}
\begin{table*}
\caption{Summary of the fitted spin values}
\centering\begin{tabular}{lccccccc}
\hline
Free or fixed &Model & {\tt kerrbb} & {\tt kerrbb2} & {\tt slimbh} & {\tt slimbh} & {\tt bhspec}& warm \\
&Type &free $f_{\rm col}$ & fitted $f_{\rm col}$ & free $f_{\rm col}$ & Atmosphere & Atmosphere &free $f_{\rm col}$ \\
\hline
Free $M_1,\,D,\,i$& $a_*$& $-0.53^{+0.34}_{-0.47}$  & $-1^{+0.05}$  & $0^{+0.22}$ & $0.71^{+0.05}_{-0.04}$   & $0.40_{-0.11}^{+0.08}$ & $0^{+0.79}$ \\
& $\chi_\nu^2$  & 833/808 & 820/810 & 815/807 & 765/812& 790/810 &751/806 \\
\hline
Fixed $M_1=10\msun$, & $a_*$& $0.39_{-0.17}^{+0.07}$  & $0.70_{-0.01}^{+0.01}$  & $0.29^{+0.03}_{-0.01}$ & $0.70^{+0.01}_{-0.01}$   & $0.60_{-0.05}^{+0.04}$ & $0.20_{-0.20}^{+0.64}$ \\
$D=10$ kpc, $i=30\degr$ & $\chi_\nu^2$  & 851/811 & 957/813 & 832/810 & 802/814& 813/812 &758/809 \\
\hline
\end{tabular}
\label{review}
\end{table*}

\subsection{Models with {\tt slimbh} and {\tt bhspec}}
\label{slimbh}

\begin{figure}
\centerline{\includegraphics[width=\columnwidth]{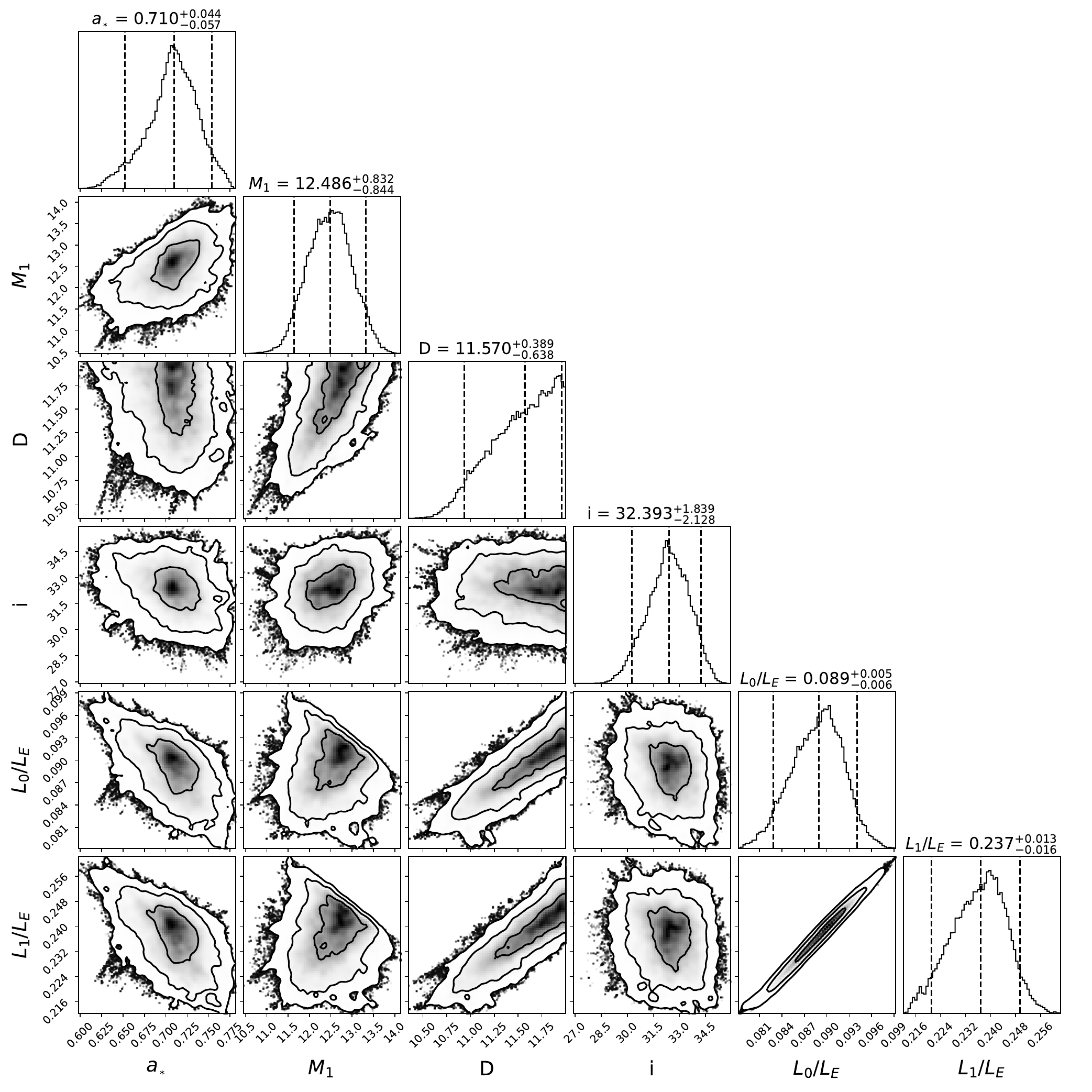}}
\caption{Correlations between the main parameters calculated by MCMC for model 3. The median values and the 90\% uncertainties are shown by the middle and surrounding dashed lines, respectively. The corresponding numerical values are given by the posterior distributions, which agree well with those in Table \ref{fits} ($\chi^2$-based).
}\label{mcmc}
\end{figure}

It is then very interesting to test other models. Indeed, the considered states of \source are relatively bright, with the luminosity being a substantial fraction of the Eddington value. At such luminosities, the accretion disk is no longer thin, which is assumed in the {\tt kerrbb} models. Thus, we model the disk using {\tt slimbh} for the viscosity parameter of $\alpha=0.1$. That model is available for $0\leq a_*\leq 0.999$, and it allows for using either a free color correction or the atmospheric calculations. For the former, $\chi_\nu^2= 329/332$, 478/467, $a_*=0.66_{-0.39}^{+0.22}$ and $0.0^{+0.24}$ for the two data sets, respectively. For the joint fit, we obtain $\chi_\nu^2= 815/807$, $a_*=0^{+0.22}$, $M_1\approx 6.7_{-0.2}^{+0.1}\msun$, $D=8.1^{+0.5}_{-0.1}$ kpc, $i=28\degr_{-3}^{+2}$, see also Table \ref{review}. The fit at $a_*=0$ with a small uncertainty indicates that a negative spin is possibly a true solution in this case, similarly to the results above. 

However, significantly positive spins are found when using {\tt slimbh} with atmospheric spectra. We obtain $\chi_\nu^2= 319/333$, 436/468, $a_*= 0.77_{-0.10}^{+0.09}$ and $0.62_{-0.14}^{+0.05}$, $M_1=13.9_{-2.5}^{+2.5}$, $11.5_{-0.8}^{+0.9} \msun$, $D=12.0_{-5.2}$, $11.8_{-1.3}^{+0.2}$ kpc, $i=32\degr_{-3}^{+4}$, $33\degr_{-2}^{+2}\degr$, for the two data sets, respectively. We have found that these two parameters are most similar to each other among the considered models. The joint fit is very good, at $\chi_\nu^2= 765/812$, with $a_*= 0.71_{-0.04}^{+0.05}$, see Table \ref{fits}, where it is denoted as model 3. The $\chi^2$ dependence on the decreasing $a_*$ is monotonic, with $\Delta\chi^2=+175$ at $a_*=0$. Thus, while {\tt slimbh} does not allow for $a_*<0$, the existence of a local minimum at those values appears unlikely (though not proven). The spectra and residuals are shown in Figure \ref{f_models}. 

An important feature of these spectra is that the scattered disk emission, which is both emitted outside and impinges on the disk, has the shape very different from a power law, especially for the 2020 data. This argues against using the model of Equation (\ref{relxill}), where the reflection component is calculated for a power law with a cutoff. When we considered that model, we obtained $\chi_\nu^2= 785/808$ and with $a_*= 0.66_{-0.02}^{+0.08}$. While the fit is worse than for the self-consistent model, Equation (\ref{xilconv}), the fitted spin is relatively similar (as well as $M_1$ and $D$). However, we find that the obtained reflection component does not correspond to the incident spectrum, which is due to scattering of the disk emission. The fitted slopes of the high-energy tails of the Comptonization emission correspond to the photon index of $\Gamma\lesssim 2$, whereas the indices of the {\tt relxillCp} component are $\Gamma\approx 3.2$ and 3.1 for the 2020 and 2021 data, respectively. Thus, while this model can fit the data, it does not capture the underlying physics, as also discussed in Section \ref{models}. This reiterates the necessity of using self-consistent models, in which scattering of the disk emission is connected to both direct emission and reflection. 

We show correlations between the main parameters for this model in Figure \ref{mcmc}. We used the Markov Chain Monte Carlo (MCMC) method \citep{Foreman-Mackey13}, as implemented in {\sc xspec}. The spin is, obviously, correlated with the mass. It is also anti-correlated with the Eddington ratios, which is due for the latter to decrease with the increasing mass. The two ratios are linearly correlated. Overall, we see that the parameters are well constrained.

We use this model to test some of our assumptions. When we assume the slab geometry in {\tt comppsc}, we find very similar parameters. We obtain $\chi_\nu^2= 767/812$, with $a_*= 0.76_{-0.03}^{+0.03}$, and the parameters of the Comptonizing coronae of $kT_0=kT_1= 14_{-1}^{+1}$ keV, $y_0=0.04_{-0.01}^{+0.01}$, $y_1=0.06_{-0.01}^{+0.01}$. Thus, our use of the fast Comptonization method in spherical geometry has only a minor effect on our results. Next, we check for the necessity of allowing the O and Fe abundances in the ISM to be different from solar. When we fix them at the solar value, the fit becomes much worse, with $\chi_\nu^2=1374/814$, confirming our assumption.

For further comparison, we consider the atmospheric thin-disk model {\tt bhspec}. We assume $\alpha=0.1$, for which the spectra are available for $0\leq a_*\leq 0.8$ only. We obtain $\chi_\nu^2= 344/333$, 452/468, $a_*=0.80_{-0.14}$ and $0.20_{-0.10}^{+0.13}$ for the two data sets, respectively. The joint fit yields $\chi_\nu^2= 790/810$, $a_*= 0.40_{-0.11}^{+0.08}$, $M_1= 9.9_{-1.1}^{+1.2}$, $D=11.4_{-1.0}^{+0.6}$ kpc, $i=33_{-4}^{+5}\degr$, see also Table \ref{review}.

Thus, we see that both models utilizing atmospheric disk spectra yield the spins positive and well above zero. The values for the joint fits are in the range of $a_*\approx 0.3$--0.8, while models using color corrections yield negative (or null, when a model does not allow for $a_*<0$) spins. Our results thus show a very strong sensitivity of the fitted spin to the assumption about the departures of the local spectra from blackbodies. 

\subsection{Warm coronae}
\label{warm_c}

We then consider the model with a warm corona, Equation (\ref{warm}). The variant of {\tt slimbh} with a free color correction should be used since the disk is now covered by an optically-thick warm scattering layer, implying that its top layers are no longer a disk atmosphere. For each of the data sets, we obtain unconstrained values of the spin, $a_*\approx 0.6_{-0.6}^{+0.4}$, and of the distance, within the assumed range of 8--12 kpc. The $\chi_\nu^2= 321/329$, 430/465 for the two data sets, respectively. The joint fit is the best among the considered models, with $\chi_\nu^2= 751/805$, where $a_*= 0^{+0.79}$, see Table \ref{fits}, where it is denoted as model 4. We see that its parameters are only weakly constrained, but the spin is consistent with being low. The F-test gives the probability the fit improvement with respect to model 3 is by chance is about 2\%. 

\subsection{Model comparison for given $M_1$, $D$ and $i$}
\label{comparison}

Above, we fitted different models to the same data, but allowing for free fitted values of $M_1$, $D$, and $i$. We now test the models for a fixed set of these parameters, $M_1=10\msun$, $D=10$ kpc, $i=30\degr$, by fitting the joint data. The results are shown in Table \ref{review}. We see the fitted values of $a_*$ span the range of $\approx 0.3$--0.7, except for the warm corona model, where it is low at the best fit but only weakly constrained. We also see that the values of $\chi^2$ are now substantially larger (except for the last model) than those with free $M_1$, $D$ and $i$, indicating that the respective models strongly prefer different values of $M_1$ and $D$.

That range of $a_*$ reflects the different treatments of the spectra in different models, as well the inclusion of the finite disk-thickness by the {\tt slimbh} model. The fits with both {\tt kerrbb} and {\tt slimbh} with free color correction give $f_{\rm col}\sim 1.9$ and $a_*\approx 0.3$--0.4, with $\Delta \chi^2= +13$ and +50, respectively, when $a_*$ is fixed at 0.7, found for the atmospheric {\tt slimbh} model and for {\tt kerrbb2}. This indicates that the effective color correction in the atmospheric models is $<1.9$. Then, the fit with {\tt kerrbb2} has $\chi^2$ much higher than the other models. On the other hand, the model with a warm corona is only weakly constrained, spanning the range of $a_*\approx 0.0$--0.7. 

\section{Discussion}
\label{discussion}

\subsection{Connections between the parameters}
\label{connection}

We have studied two sets of the X-ray spectra of \source in the soft state. They are strongly dominated by the apparent disk blackbody components. Fitting the {\tt diskbb} model yields the normalization, $N_{\rm dbb}$, compatible with being the same for both data sets. Since $N_{\rm dbb}$ determines the radius (see below), this argues for the disk inner radius being the same in both cases. Since, in turn, the ISCO gives the only radius constant in the disk theory, it is likely to be at the ISCO. Also, similar values of $N_{\rm dbb}$ were obtained in a number of other studies of the soft state of \source. For example, \citet{Plant14} obtained $N_{\tt dbb}=3249\pm 194$ as an average for the soft state of three outbursts, of 2002, 2004 and 2007. Then, \citet{Sridhar20} obtained $N_{\tt dbb}=3810_{-110}^{+90}$ and $3620_{-320}^{+330}$ for the softest spectra from the 2002 and 2004 outbursts, respectively. The somewhat larger values than those obtained in our analysis appear to be due to the flux calibrations of the PCA detector onboard Rossi X-ray Timing Explorer used by those authors being higher than that of NuSTAR (used by us), see \citep{Madsen17}. Also, those studies used different spectral models (with the power law and reflection components being separate from the disk emission) than that in our work.

The distance, inclination and the inner radius are related to $N_{\rm dbb}$ by
\begin{equation}
R_{\rm in}=10^5 f_{\rm in} f_{\rm col}^2 \frac{D}{10\,{\rm kpc}} \left(\frac{N_{\tt dbb}}{\cos i}\right)^{1/2}\,{\rm cm},
\label{rin_disk}
\end{equation}
where $f_{\rm in}<1$ is a correction factor accounting for {\tt diskbb} not including the zero-stress term at the ISCO. \citet{Kubota98} estimated it as $\approx 0.41$ for the case of $a_*=0$. For $N_{\tt dbb}=3000$ and $i=32\degr$, the inner radius in units of the gravitational radius, $R_{\rm g}\equiv G M_1/c^2$, becomes
\begin{equation}
\frac{R_{\rm in}}{R_{\rm g}}\approx 4.8 \frac{D}{10\,{\rm kpc}} \left(\frac{M_1}{10\msun}\right)^{-1} \frac{f_{\rm in}}{0.41} \left(\frac{f_{\rm col}}{1.7}\right)^2,
\label{rin_solution}
\end{equation}
which indicates the BH is slowly rotating, or counter-rotating for high values of $D/M_1$. Thus, a given value of $D/M_1$ would give us an estimate of the spin, though modulo the values of $f_{\rm in}$ and $f_{\rm col}$ and neglecting relativistic effects. Still, we can compare values implied by Equation (\ref{rin_solution}) with those fitted with different models. For models 2, 3, 4 in Table \ref{fits} assuming the default values of $f_{\rm in}$ and $f_{\rm col}$, we obtain $R_{\rm in}/R_{\rm g}\approx 7.3$, 4.4, 8.1, giving the spins of $a_*\approx -0.42$, 0.46 and $-0.69$, respectively. For models 2 and 3, these estimates are in fair agreement with the fitted values of $a_*$. The value for model 4 may indicate that the true spin for this model is negative. However, including $f_{\rm col}\approx 1.4$ (Table \ref{fits}), gives $a_*\approx 0.15$, within the uncertainties of the fit.

Then, the fact that a given model yields specific values of $M_1$ and $D$ follows from the relativistic effects, the dependence of the correction for the zero-stress boundary at the ISCO on the spin, finite disk-thickness, and the handling of the modifications to the local blackbody spectra. 

\subsection{The mass function and preferred models}
\label{preferred}

Comparing models 2 and 3, we see that the data strongly prefer the latter, which takes into account the atmospheric spectra and the disk finite thickness. Relatively similar results, though with a worse $\chi^2$, were obtained with the thin-disk atmospheric model {\tt bhspec} (Section \ref{slimbh}), which points to the major importance of the atmospheric effects. Then, model 3 yields relatively large both the mass and the distance, $M_1\approx 12$--$13\msun$, and $D\approx 11$--12 kpc. 

We note that the BH mass, $M_1$, is connected to the binary inclination, $i_{\rm b}$, by the mass function and the mass ratio, determined by \citet{Heida17} as
\begin{equation}
\frac{M_1 \sin^3 i_{\rm b}}{(1+M_2/M_1)^2}=1.91\pm 0.08\msun,\quad \frac{M_2}{M_1}=0.18\pm 0.05,
\label{mass}
\end{equation}
where $M_2$ is the donor mass. Our joint fits yield $i\approx 30\degr$--$34\degr$ in all cases listed in Table \ref{fits}. If $i_{\rm b}=i$ in the above range, we have $13\msun\lesssim M_1\lesssim 24\msun$, anticorrelated with $i$, and including the uncertainties of the mass function and the mass ratio. We see in Table \ref{fits} and Section \ref{results} that models 3 and 4 have the ranges of the BH mass compatible with Equation (\ref{mass}). In the case of model 3, with the atmospheric version of {\tt slimbh}, the fit and the parameters remain almost the same as before (assuming the lowest mass allowed by Equation \ref{mass}); $\chi^2_\nu=769/813$, $a_*=0.70^{+0.03}_{-0.03}$, $D=11.8_{-0.7}^{+0.2}$ kpc, $i=34^{+1}_{-1}\degr$, and $M=13.2\msun$ at the best fit of $i$. For model 4, with a warm corona, $\chi^2_\nu=753/807$, $a_*=0.71_{-0.16}^{+0.24}$, $D=10.5_{-0.9}^{+1.5}$ kpc, $i=33_{-2}^{+1}\degr$, and $M=14.0\msun$ at the best fit of $i$. 

Our preferred model by both the $\chi^2$ criterion and the virtue of Ockham's razor, i.e., allowing for the simplest case of $i=i_{\rm b}$ and no warm corona, is model 3. This model has relatively high spin, mass and the distance. Still, an inner part of the disk, which determines the observed X-rays, can be warped, in which case we would have $i_{\rm b}>i$, where the $>$ sign is implied by the low masses fitted in other models. The warm corona model remains viable as well, but, as we see above, it is no longer compatible with both the mass function and a low spin.

We would be able to constrain the spin better and to select the best model better if there were better estimates of the mass and the distance. Hopefully, they will become available in the future. 

\subsection{Outstanding issues}
\label{caveats}

As pointed out above, our preferred model is that using the atmospheric spectra of \citet{Davis05} and \citet{Davis06}. An important caveat here is that it uses the disk model \citep{SS73, NT73}, which unambiguously predicts the disk to be both viscously and thermally unstable when dominated by radiation pressure \citep{Lightman74, Shakura76}. For $M_1\sim 10 \msun$, this occurs for $L/L_{\rm E}\gtrsim 0.02$ (e.g., \citealt{SZ94}), which limit is way below the values of $L/L_{\rm E}$ for our data, which then predicts the disk in our observations to be strongly unstable. However, the disk in our cases is extremely stable, with the 0.5--3 keV rms variability of $\lesssim 0.5\%$, see Section \ref{data}. The same issue occurs for a number of other accreting BH XRBs in the soft state \citep{GD04}. The stability problem can be solved if strong toroidal magnetic field is present, e.g., \citet{Begelman07, Begelman17}. Those models, however, have the pressure dominated by the magnetic field, which is expected to substantially alter the emitted X-ray spectra (e.g., \citealt{Davis19}). Thus, we do not expect the standard model to give a good description of the soft state of BH XRBs (as discussed, e.g., in \citealt{Zdziarski24b}), while it apparently gives a very good one in the present case. 

Furthermore, a major result from gravitational wave studies is that the spins of merging binary BHs are low for most observed mergers \citep{Abbott23}. In particular, that analysis shows this is the case for the first-formed (and usually most massive) BH (fig.\ 17 in \citealt{Abbott23}), whose spin is also expected to be small by current binary evolution theory \citep{Fuller_Ma19}. Indeed, standard stellar models with efficient angular momentum transport \citep{Spruit02, Belczynski20} yield the natal spins of $a_*\sim 0.1$. Since the observed masses of donors in LXMBs are low, $\lesssim\! 1\msun$, and an increase of the BH spin to $a_*\sim 1$ requires doubling the BH mass \citep{Bardeen70}, only models in which the initial donor masses were much higher than those in the observed LMXBs can account for high BH spins \citep{Fragos15}, which case remains highly uncertain. Furthermore, the evolutionary models of \citet{Fragos15} assumed fully conservative accretion and thus neglected outflows even during hyper-Eddington phases, which is another uncertainty, see, e.g., \citet{Poutanen07}. 

If the accretion is indeed insufficient to spin up BHs in LMXBs, a viable alternative is provided by the model with a warm corona. While model 4 is not strongly required based on the F-test, it is neither ruled out. It is compatible with low spins, fully consistent with the values measured in merging BHs. Similarly, the three known high-mass BH XRBs (see Section \ref{models}) were claimed to have very spins. While the donors have high masses in those systems, their short lifetime prevents a significant spin up unless the accretion onto the BH is highly super-Eddington, see e.g., a discussion for Cyg X-1 in \citet{Zdziarski24b}. However, their spectrally-measured spins become low if warm coronae cover their accretion disks \citep{Belczynski24, Zdziarski24a, Zdziarski24b}, which removes the need for highly super-Eddington accretion.

In our modeling, we have implicitly assumed that the disk surrounding the BH is aligned along the BH equatorial plane. If the BH spin is misaligned with respect to the binary axis, the models used by us are no more strictly valid. Another issue regards the emission from below ISCO (the plunging region). Current studies, e.g., \citet{Mummery24} predict substantial emission from that region in the form of a soft/steep high-energy tail beyond the disk blackbody, and increasing with the decreasing spin, thus strongest at $a_*=-1$. We see no such component in our spectra. We also notice that at present there is no consensus regarding the physical description of that process \citep{Lasota24}.

Another effect we have neglected is the reflection of the blackbody radiation returning to the disk\footnote{Note that the option of {\tt kerrbb} and {\tt kerrbb2} to switch on self-irradiation assumes the disk is completely absorbing, which then only slightly increases the local temperatures.} due to light bending \citep{Schnittman09, Mirzaev24}. This effect forms of a weak and soft high-energy tail beyond the disk blackbody with its amplitude increasing with the spin. It becomes noticeable only at $a_*\gtrsim 0.9$ \citep{Schnittman09}, which range falls outside those found in this work. Still, we see no such component in our spectra. 

\subsection{Negative spins}
\label{negative}

With the widely used models {\tt kerrbb} and {\tt kerrbb2}, we have obtained strongly negative spins for both individual and joint fits with high statistical significance. While the values of the $\chi^2$ of those fits are higher than those for our preferred model ({\tt slimbh} with the atmospheric spectra), a reliable distinction between them would require accurate knowledge of $M_1$ and $D$, which is presently not available. Thus, we should take this possibility seriously. 

While retrograde accretion onto a BH in a binary is in principle possible, it requires either a a BH spin reversal after its formation or a formation of the binary by dynamical interactions. The former can happen only in rare cases, while the latter requires the binary to be in a globular cluster, which is not the case for \source. Still, the system could have been formed in a globular cluster and then ejected. 

Indeed, an important result from the gravitational wave studies is that the distribution of the effective spin (which is a weighted sum of the individual spins projected onto the orbital axis) has a highly significant negative part \citep{Abbott23}. As noted by \citet{Tauris22}, a substantial part of those spins had to be from systems produced in isolated binaries, which is not explained by current stellar evolution theories. He proposed that the misaligned and anti-aligned systems were produced during so-called 'spin tossing' during the core collapse. This then implies that a fraction of the existing LMXBs should have negative BH spin. 

There have been several sources for which a negative spin was claimed in the past. \citet{Reis13} studied Swift J1910.2--0546 (also known as MAXI J1910--057), a transient BH LMXB. The binary was in the soft intermediate state. They found negative values of the spin using the reflection spectroscopy assuming the incident spectrum is a power law. This is a relatively unreliable method, especially in soft states, where the scattered spectrum incident on the disk can have a spectrum different from a power law (as is the case for our study, see the green curves in Figure \ref{f_models}). Also, as the authors mentioned, an alternative explanation for the relatively large disk inner radius they found can be disk truncation.

\citet{Morningstar14} studied the soft state of the transient BH LXMB GS 1124--683 (also known as Nova Muscae 1991). They determined the spin using {\tt kerrbb} as $a_*= -0.25_{-0.64}^{+0.05}$ based on a determination of the BH mass and the distance of $7.24\pm 0.70\msun$, $5.89\pm 0.25$ kpc, respectively. However, both values were revised by \citet{Wu16} to $11.0_{-1.4}^{+2.1}\msun$ and $4.95^{+0.69}_{-0.65}$ kpc, respectively. An increase of the mass and a decrease of the distance both led to a reduction of the inferred disk inner radius, and, in turn, to an increase of the spin. Indeed, \citet{Chen16} fitted similar data as \citet{Morningstar14} using {\tt kerrbb2} obtaining $a_*= 0.63_{-0.19}^{+0.16}$.

\citet{Middleton14} studied the microquasar XMMU J004243.6+412519 in M31 in the soft spectral state during its outburst in 2012. The distance to M31 is well known, and the BH mass was estimated at $10\msun$ based on the evidence of the accretion rate was close to the Eddington limit at the peak X-ray brightness. They fitted the spectra using a version of atmospheric code {\tt bhspec} covering both positive and negative spins and for $\alpha=0.01$. They found clearly negative spin values at the best fits, $a_*= -1$ up to $\sim\! -0.4$ depending on the uncertain inclination. The main uncertainty appears to be the BH mass. If it were, e.g., $15\msun$ instead, the obtained spin values would be positive. The authors did not test other available disk models. However, a result of the present work is that models with color corrections, e.g., {\tt kerrbb} or {\tt kerrbb2} give lower spin values than {\tt bhspec} (Section \ref{comparison}). Thus, we would not expect a higher value of $a_*$ if the former models were used.

\citet{Rout20} studied an X-ray spectrum of MAXI J1659--152 in the rising phase of its outburst. That LMXB was then in a luminous hard-intermediate state. The spectrum consisted of components due to disk blackbody, Comptonization and reflection. In their preferred model, the disk contributed only 20\% of the bolometric flux of $\approx 1.7\times 10^{-8}$ erg cm$^{-2}$ s$^{-1}$, and the reflection contribution was very weak. The Comptonization power law was relatively hard, with the photon index of $\Gamma \approx 1.9$. For a fiducial distance of 6 kpc and a BH mass of $10\msun$, $L/L_{\rm E}\approx 0.06$. The authors based their claim of $a_*\approx -1$ on the fitted inner disk radius of about $10 R_{\rm g}$ and some arguments on the accretion efficiency (unclear to us). However, in our opinion it is quite possible that the spin is prograde and the accretion disk is truncated, as likely to occur in the hard or hard intermediate state, see, e.g., \citet{DGK07}. This is supported by the strong dominance of the Comptonization component, likely formed by a hot plasma flow below the disk truncation radius. 

Summarizing those claims, we find the result of \citet{Middleton14} as the only relatively reliable one. Still, even their determination of the negative spin could be revised upward if the BH in XMMU J004243.6+412519 is more massive than their assumed value. Thus, there seems to be no BH LMXB with confirmed retrograde accretion as yet.

\section{Conclusions}
\label{conclusions}

Our major result is a confirmation of strong model dependence of the measured values of the black-hole spin \citep{Belczynski24,Zdziarski24a, Zdziarski24b}, now using different models of relativistic disks and for an LMXB. The fitted spin strongly depends on the chosen model for the cases either allowing free mass and distance or setting them fixed. We have studied two data sets of outstanding quality of \source in the soft state, and have found that we are able to constrain the mass, spin, distance and inclination for each of the models (except for the warm corona). However, those fitted values significantly differ among the considered models, see Table \ref{review} for the spins, with all of of the fits appearing reasonable with $\chi^2_\nu\sim 1$. Only the inclination was found well constrained within $30\degr$--$34\degr$ for all of the models. 

Specifically, we have found strongly negative spins and low BH masses with the widely used thin-disk models {\tt{kerrbb}} and {\tt{kerrbb2}}. Those models employ color corrections to treat the atmospheric departures from the blackbody emission, though the latter uses those corrections fitted to the atmospheric calculations. On the other hand, we obtained relatively large positive spins and high masses with the two models using directly the atmospheric spectra \citep{Davis05, Davis06}, namely {\tt slimbh} and {\tt bhspec}. Also, the quality of these fits was much better than those using color corrections. When adding a warm corona above the disk, we have found that the model parameters are weakly constrained, but the spin is consistent with zero. Furthermore, we have found that even when fixing the mass, distance and inclination, different models give significantly different spin parameters, $\Delta a_*\sim 0.2$--0.3, see Table \ref{review}.

We then considered the mass function, known for this object. It implies relatively large masses if the fitted inclination equals the binary inclination. This would be the case when the binary axis and the spin axis are aligned, which is not certain. We have found that only models with either the atmospheric calculations or warm coronae satisfy the mass function constraint for the fitted range of the inclination.

Our model preferred by the $\chi^2$ value, the agreement of the fitted mass of $12.5^{+1.0}_{-0.3}\msun$ with the mass function, and the relative simplicity is model 3 in Table \ref{fits}, which includes the slim disk model {\tt slimbh} in the version using the atmospheric calculations. It yields a relatively large spin of $a_*=0.71_{-0.04}^{+0.05}$, the inclination of $i=32\degr \pm 2\degr$ and the distance of $11.5^{+0.5}_{-0.8}$ kpc. 

This model assumes the standard vertical support by the MRI turbulence and it is thus subject to the viscous and thermal instabilities. However, we have found that the disk is almost completely stable for both of the studied measurements, which remains unexplained. Furthermore, it is unclear how the large spin would be achieved in the system. The gravitational wave results show that the natal BH spins are low, and the current spin would have to be from accretion. In that case, the initial donor mass would have be at least $\sim\! 10\msun$. 

The large spin problem can be solved when adopting our model with a warm corona (see Table \ref{fits}), which is compatible with a low spin. Unfortunately, the additional degrees of freedom of this model make its parameters only weakly constrained, e.g., $a_*=0^{+0.79}$. 

We have also reviewed past findings of negative BH spin in LMXBs, and found none of them fully convincing. On the other hand, the merger results show a significant fraction of weakly negative spins, and, according to the model of \citet{Tauris22}, the spin of the BH could be reversed during the core collapse. 

Finally, we have shown that the spectra from Comptonization incident on the disk have the shape far from a power law with a cutoff. This shows the necessity of using self-consistent reflection models in the soft spectral state. 

\section*{Acknowledgements}
We thank Michal Bursa and Odele Straub for consultations and help with using {\tt slimbh}, and Thomas Tauris and the referee for valuable comments. We acknowledge support from the Polish National Science Center grants 2019/35/B/ST9/03944 and 2023/48/Q/ST9/00138. MS acknowledges support from the National Science Center grant 2023/50/A/ST9/00527.

\bibliographystyle{aasjournal}
\bibliography{../../allbib} 

\begin{thebibliography}{}
\expandafter\ifx\csname natexlab\endcsname\relax\def\natexlab#1{#1}\fi
\providecommand{\url}[1]{\href{#1}{#1}}
\providecommand{\dodoi}[1]{doi:~\href{http://doi.org/#1}{\nolinkurl{#1}}}
\providecommand{\doeprint}[1]{\href{http://ascl.net/#1}{\nolinkurl{http://ascl.net/#1}}}
\providecommand{\doarXiv}[1]{\href{https://arxiv.org/abs/#1}{\nolinkurl{https://arxiv.org/abs/#1}}}

\bibitem[{{Abbott} {et~al.}(2023){Abbott}, {Abbott}, {Acernese}, {Ackley},
  {Adams}, {Adhikari}, {Adhikari}, {Adya}, {Affeldt}, {Agarwal}, \&
  et~al.}]{Abbott23}
{Abbott}, R., {Abbott}, T.~D., {Acernese}, F., {et~al.} 2023, Physical Review
  X, 13, 011048, \dodoi{10.1103/PhysRevX.13.011048}

\bibitem[{{Abramowicz} {et~al.}(1988){Abramowicz}, {Czerny}, {Lasota}, \&
  {Szuszkiewicz}}]{Abramowicz88}
{Abramowicz}, M.~A., {Czerny}, B., {Lasota}, J.~P., \& {Szuszkiewicz}, E. 1988,
  \apj, 332, 646, \dodoi{10.1086/166683}

\bibitem[{{Anders} \& {Grevesse}(1989)}]{AG89}
{Anders}, E., \& {Grevesse}, N. 1989, \gca, 53, 197,
  \dodoi{10.1016/0016-7037(89)90286-X}

\bibitem[{{Arnaud}(1996)}]{Arnaud96}
{Arnaud}, K.~A. 1996, ASP Conference Series, Vol. 101, {XSPEC: The First Ten
  Years}, ed. G.~H. {Jacoby} \& J.~{Barnes} (ASP), 17

\bibitem[{{Ballantyne} {et~al.}(2024){Ballantyne}, {Sudhakar}, {Fairfax},
  {Bianchi}, {Czerny}, {De Rosa}, {De Marco}, {Middei}, {Palit}, {Petrucci},
  {R{\'o}{\.z}a{\'n}ska}, \& {Ursini}}]{Ballantyne24}
{Ballantyne}, D.~R., {Sudhakar}, V., {Fairfax}, D., {et~al.} 2024, \mnras, 530,
  1603, \dodoi{10.1093/mnras/stae944}

\bibitem[{{Bambi} {et~al.}(2021){Bambi}, {Brenneman}, {Dauser}, {Garcia},
  {Grinberg}, {Ingram}, {Jiang}, {Kara}, {Liu}, {Lohfink}, {Marinucci},
  {Mastroserio}, {Middei}, {Nampalliwar}, {Niedzwiecki}, {Steiner}, {Tripathi},
  \& {Zdziarski}}]{Bambi21}
{Bambi}, C., {Brenneman}, L.~W., {Dauser}, T., {et~al.} 2021, \ssr, 217, 65,
  \dodoi{doi.org/10.1007/s11214-021-00841-8}

\bibitem[{{Bardeen}(1970)}]{Bardeen70}
{Bardeen}, J.~M. 1970, \nat, 226, 64, \dodoi{10.1038/226064a0}

\bibitem[{{Bardeen} {et~al.}(1972){Bardeen}, {Press}, \&
  {Teukolsky}}]{Bardeen72}
{Bardeen}, J.~M., {Press}, W.~H., \& {Teukolsky}, S.~A. 1972, \apj, 178, 347,
  \dodoi{10.1086/151796}

\bibitem[{{Basak} \& {Zdziarski}(2016)}]{Basak16}
{Basak}, R., \& {Zdziarski}, A.~A. 2016, \mnras, 458, 2199,
  \dodoi{10.1093/mnras/stw420}

\bibitem[{{Begelman} \& {Pringle}(2007)}]{Begelman07}
{Begelman}, M.~C., \& {Pringle}, J.~E. 2007, \mnras, 375, 1070,
  \dodoi{10.1111/j.1365-2966.2006.11372.x}

\bibitem[{{Begelman} \& {Silk}(2017)}]{Begelman17}
{Begelman}, M.~C., \& {Silk}, J. 2017, \mnras, 464, 2311,
  \dodoi{10.1093/mnras/stw2533}

\bibitem[{{Belczy{\'n}ski} {et~al.}(2024){Belczy{\'n}ski}, {Done}, {Hagen},
  {Lasota}, \& {Sen}}]{Belczynski24}
{Belczy{\'n}ski}, K., {Done}, C., {Hagen}, S., {Lasota}, J.-P., \& {Sen}, K.
  2024, \aap, 690, A21, \dodoi{10.1051/0004-6361/202450229}

\bibitem[{{Belczy{\'n}ski} {et~al.}(2020){Belczy{\'n}ski}, {Klencki}, {Fields},
  {Olejak}, {Berti}, {Meynet}, {Fryer}, {Holz}, {O'Shaughnessy}, {Brown},
  {Bulik}, {Leung}, {Nomoto}, {Madau}, {Hirschi}, {Kaiser}, {Jones}, {Mondal},
  {Chruslinska}, {Drozda}, {Gerosa}, {Doctor}, {Giersz}, {Ekstrom}, {Georgy},
  {Askar}, {Baibhav}, {Wysocki}, {Natan}, {Farr}, {Wiktorowicz}, {Coleman
  Miller}, {Farr}, \& {Lasota}}]{Belczynski20}
{Belczy{\'n}ski}, K., {Klencki}, J., {Fields}, C.~E., {et~al.} 2020, \aap, 636,
  A104, \dodoi{10.1051/0004-6361/201936528}

\bibitem[{{Chen} {et~al.}(2016){Chen}, {Gou}, {McClintock}, {Steiner}, {Wu},
  {Xu}, {Orosz}, \& {Xiang}}]{Chen16}
{Chen}, Z., {Gou}, L., {McClintock}, J.~E., {et~al.} 2016, \apj, 825, 45,
  \dodoi{10.3847/0004-637X/825/1/45}

\bibitem[{{Dauser} {et~al.}(2010){Dauser}, {Wilms}, {Reynolds}, \&
  {Brenneman}}]{Dauser10}
{Dauser}, T., {Wilms}, J., {Reynolds}, C.~S., \& {Brenneman}, L.~W. 2010,
  \mnras, 409, 1534, \dodoi{10.1111/j.1365-2966.2010.17393.x}

\bibitem[{{Davis} {et~al.}(2005){Davis}, {Blaes}, {Hubeny}, \&
  {Turner}}]{Davis05}
{Davis}, S.~W., {Blaes}, O.~M., {Hubeny}, I., \& {Turner}, N.~J. 2005, \apj,
  621, 372, \dodoi{10.1086/427278}

\bibitem[{{Davis} \& {El-Abd}(2019)}]{Davis19}
{Davis}, S.~W., \& {El-Abd}, S. 2019, \apj, 874, 23,
  \dodoi{10.3847/1538-4357/ab05c5}

\bibitem[{{Davis} \& {Hubeny}(2006)}]{Davis06}
{Davis}, S.~W., \& {Hubeny}, I. 2006, \apjs, 164, 530, \dodoi{10.1086/503549}

\bibitem[{{De Marco} {et~al.}(2015){De Marco}, {Ponti}, {Mu{\~n}oz-Darias}, \&
  {Nandra}}]{DeMarco15}
{De Marco}, B., {Ponti}, G., {Mu{\~n}oz-Darias}, T., \& {Nandra}, K. 2015,
  \apj, 814, 50, \dodoi{10.1088/0004-637X/814/1/50}

\bibitem[{{Done} {et~al.}(2007){Done}, {Gierli{\'n}ski}, \& {Kubota}}]{DGK07}
{Done}, C., {Gierli{\'n}ski}, M., \& {Kubota}, A. 2007, \aapr, 15, 1,
  \dodoi{10.1007/s00159-007-0006-1}

\bibitem[{{Dzie{\l}ak} {et~al.}(2019){Dzie{\l}ak}, {Zdziarski}, {Szanecki}, {De
  Marco}, {Nied{\'z}wiecki}, \& {Markowitz}}]{Dzielak19}
{Dzie{\l}ak}, M.~A., {Zdziarski}, A.~A., {Szanecki}, M., {et~al.} 2019, \mnras,
  485, 3845, \dodoi{10.1093/mnras/stz668}

\bibitem[{{Foreman-Mackey} {et~al.}(2013){Foreman-Mackey}, {Hogg}, {Lang}, \&
  {Goodman}}]{Foreman-Mackey13}
{Foreman-Mackey}, D., {Hogg}, D.~W., {Lang}, D., \& {Goodman}, J. 2013, \pasp,
  125, 306, \dodoi{10.1086/670067}

\bibitem[{{Fragos} \& {McClintock}(2015)}]{Fragos15}
{Fragos}, T., \& {McClintock}, J.~E. 2015, \apj, 800, 17,
  \dodoi{10.1088/0004-637X/800/1/17}

\bibitem[{{Fuller} \& {Ma}(2019)}]{Fuller_Ma19}
{Fuller}, J., \& {Ma}, L. 2019, \apjl, 881, L1,
  \dodoi{10.3847/2041-8213/ab339b}

\bibitem[{{Garc{\'{\i}}a} {et~al.}(2013){Garc{\'{\i}}a}, {Dauser}, {Reynolds},
  {Kallman}, {McClintock}, {Wilms}, \& {Eikmann}}]{Garcia13}
{Garc{\'{\i}}a}, J.~A., {Dauser}, T., {Reynolds}, C.~S., {et~al.} 2013, \apj,
  768, 146, \dodoi{10.1088/0004-637X/768/2/146}

\bibitem[{{Garc{\'{\i}}a} {et~al.}(2015){Garc{\'{\i}}a}, {Steiner},
  {McClintock}, {Remillard}, {Grinberg}, \& {Dauser}}]{Garcia15}
{Garc{\'{\i}}a}, J.~A., {Steiner}, J.~F., {McClintock}, J.~E., {et~al.} 2015,
  \apj, 813, 84, \dodoi{10.1088/0004-637X/813/2/84}

\bibitem[{{Garc{\'{\i}}a} {et~al.}(2018){Garc{\'{\i}}a}, {Steiner}, {Grinberg},
  {Dauser}, {Connors}, {McClintock}, {Remillard}, {Wilms}, {Harrison}, \&
  {Tomsick}}]{Garcia18}
{Garc{\'{\i}}a}, J.~A., {Steiner}, J.~F., {Grinberg}, V., {et~al.} 2018, \apj,
  864, 25, \dodoi{10.3847/1538-4357/aad231}

\bibitem[{{Gierli{\'n}ski} \& {Done}(2004)}]{GD04}
{Gierli{\'n}ski}, M., \& {Done}, C. 2004, \mnras, 347, 885,
  \dodoi{10.1111/j.1365-2966.2004.07266.x}

\bibitem[{{Hagen} {et~al.}(2024){Hagen}, {Done}, {Silverman}, {Li}, {Liu},
  {Ren}, {Buchner}, {Merloni}, {Nagao}, \& {Salvato}}]{Hagen24}
{Hagen}, S., {Done}, C., {Silverman}, J.~D., {et~al.} 2024, \mnras, 534, 2803,
  \dodoi{10.1093/mnras/stae2272}

\bibitem[{{Heida} {et~al.}(2017){Heida}, {Jonker}, {Torres}, \&
  {Chiavassa}}]{Heida17}
{Heida}, M., {Jonker}, P.~G., {Torres}, M.~A.~P., \& {Chiavassa}, A. 2017,
  \apj, 846, 132, \dodoi{10.3847/1538-4357/aa85df}

\bibitem[{{Kaastra} \& {Bleeker}(2016)}]{Kaastra16}
{Kaastra}, J.~S., \& {Bleeker}, J.~A.~M. 2016, \aap, 587, A151,
  \dodoi{10.1051/0004-6361/201527395}

\bibitem[{{Kang} {et~al.}(2024){Kang}, {Done}, {Hagen}, {Temple}, {Silverman},
  {Li}, \& {Liu}}]{Kang24}
{Kang}, J.-L., {Done}, C., {Hagen}, S., {et~al.} 2024, arXiv e-prints,
  arXiv:2410.06730, \dodoi{10.48550/arXiv.2410.06730}

\bibitem[{{Kolehmainen} \& {Done}(2010)}]{Kolehmainen10}
{Kolehmainen}, M., \& {Done}, C. 2010, \mnras, 406, 2206,
  \dodoi{10.1111/j.1365-2966.2010.16835.x}

\bibitem[{{Kolehmainen} {et~al.}(2011){Kolehmainen}, {Done}, \& {D{\'\i}az
  Trigo}}]{Kolehmainen11}
{Kolehmainen}, M., {Done}, C., \& {D{\'\i}az Trigo}, M. 2011, \mnras, 416, 311,
  \dodoi{10.1111/j.1365-2966.2011.19040.x}

\bibitem[{{Kubota} {et~al.}(1998){Kubota}, {Tanaka}, {Makishima}, {Ueda},
  {Dotani}, {Inoue}, \& {Yamaoka}}]{Kubota98}
{Kubota}, A., {Tanaka}, Y., {Makishima}, K., {et~al.} 1998, \pasj, 50, 667,
  \dodoi{10.1093/pasj/50.6.667}

\bibitem[{{Lampton} {et~al.}(1976){Lampton}, {Margon}, \& {Bowyer}}]{Lampton76}
{Lampton}, M., {Margon}, B., \& {Bowyer}, S. 1976, \apj, 208, 177,
  \dodoi{10.1086/154592}

\bibitem[{{Lasota} \& {Abramowicz}(2024)}]{Lasota24}
{Lasota}, J.-P., \& {Abramowicz}, M. 2024, arXiv e-prints, arXiv:2410.06200,
  \dodoi{10.48550/arXiv.2410.06200}

\bibitem[{{Li} {et~al.}(2005){Li}, {Zimmerman}, {Narayan}, \&
  {McClintock}}]{Li05}
{Li}, L.-X., {Zimmerman}, E.~R., {Narayan}, R., \& {McClintock}, J.~E. 2005,
  \apjs, 157, 335, \dodoi{10.1086/428089}

\bibitem[{{Lightman} \& {Eardley}(1974)}]{Lightman74}
{Lightman}, A.~P., \& {Eardley}, D.~M. 1974, \apjl, 187, L1,
  \dodoi{10.1086/181377}

\bibitem[{{Ludlam} {et~al.}(2015){Ludlam}, {Miller}, \& {Cackett}}]{Ludlam15}
{Ludlam}, R.~M., {Miller}, J.~M., \& {Cackett}, E.~M. 2015, \apj, 806, 262,
  \dodoi{10.1088/0004-637X/806/2/262}

\bibitem[{{Madsen} {et~al.}(2022){Madsen}, {Forster}, {Grefenstette},
  {Harrison}, \& {Miyasaka}}]{Madsen22}
{Madsen}, K.~K., {Forster}, K., {Grefenstette}, B., {Harrison}, F.~A., \&
  {Miyasaka}, H. 2022, Journal of Astronomical Telescopes, Instruments, and
  Systems, 8, 034003, \dodoi{10.1117/1.JATIS.8.3.034003}

\bibitem[{{Madsen} {et~al.}(2017){Madsen}, {Forster}, {Grefenstette},
  {Harrison}, \& {Stern}}]{Madsen17}
{Madsen}, K.~K., {Forster}, K., {Grefenstette}, B.~W., {Harrison}, F.~A., \&
  {Stern}, D. 2017, \apj, 841, 56, \dodoi{10.3847/1538-4357/aa6970}

\bibitem[{{Magdziarz} \& {Zdziarski}(1995)}]{MZ95}
{Magdziarz}, P., \& {Zdziarski}, A.~A. 1995, \mnras, 273, 837

\bibitem[{{Mahmoud} {et~al.}(2019){Mahmoud}, {Done}, \& {De Marco}}]{Mahmoud19}
{Mahmoud}, R.~D., {Done}, C., \& {De Marco}, B. 2019, \mnras, 486, 2137,
  \dodoi{10.1093/mnras/stz933}

\bibitem[{{McClintock} {et~al.}(2014){McClintock}, {Narayan}, \&
  {Steiner}}]{McClintock14}
{McClintock}, J.~E., {Narayan}, R., \& {Steiner}, J.~F. 2014, \ssr, 183, 295,
  \dodoi{10.1007/s11214-013-0003-9}

\bibitem[{{McClintock} {et~al.}(2006){McClintock}, {Shafee}, {Narayan},
  {Remillard}, {Davis}, \& {Li}}]{McClintock06}
{McClintock}, J.~E., {Shafee}, R., {Narayan}, R., {et~al.} 2006, \apj, 652,
  518, \dodoi{10.1086/508457}

\bibitem[{{Middleton} {et~al.}(2014){Middleton}, {Miller-Jones}, \&
  {Fender}}]{Middleton14}
{Middleton}, M.~J., {Miller-Jones}, J. C.~A., \& {Fender}, R.~P. 2014, \mnras,
  439, 1740, \dodoi{10.1093/mnras/stu056}

\bibitem[{{Miller} {et~al.}(2008){Miller}, {Reynolds}, {Fabian}, {Cackett},
  {Miniutti}, {Raymond}, {Steeghs}, {Reis}, \& {Homan}}]{Miller08}
{Miller}, J.~M., {Reynolds}, C.~S., {Fabian}, A.~C., {et~al.} 2008, \apjl, 679,
  L113, \dodoi{10.1086/589446}

\bibitem[{{Mirzaev} {et~al.}(2024){Mirzaev}, {Bambi}, {Abdikamalov}, {Jiang},
  {Liu}, {Riaz}, \& {Shashank}}]{Mirzaev24}
{Mirzaev}, T., {Bambi}, C., {Abdikamalov}, A.~B., {et~al.} 2024, \apj, 976,
  229, \dodoi{10.3847/1538-4357/ad8a63}

\bibitem[{{Mitsuda} {et~al.}(1984){Mitsuda}, {Inoue}, {Koyama}, {Makishima},
  {Matsuoka}, {Ogawara}, {Shibazaki}, {Suzuki}, {Tanaka}, \&
  {Hirano}}]{Mitsuda84}
{Mitsuda}, K., {Inoue}, H., {Koyama}, K., {et~al.} 1984, \pasj, 36, 741

\bibitem[{{Morningstar} {et~al.}(2014){Morningstar}, {Miller}, {Reis}, \&
  {Ebisawa}}]{Morningstar14}
{Morningstar}, W.~R., {Miller}, J.~M., {Reis}, R.~C., \& {Ebisawa}, K. 2014,
  \apjl, 784, L18, \dodoi{10.1088/2041-8205/784/2/L18}

\bibitem[{{Mu{\~n}oz-Darias} {et~al.}(2013){Mu{\~n}oz-Darias}, {Coriat},
  {Plant}, {Ponti}, {Fender}, \& {Dunn}}]{Munoz13}
{Mu{\~n}oz-Darias}, T., {Coriat}, M., {Plant}, D.~S., {et~al.} 2013, \mnras,
  432, 1330, \dodoi{10.1093/mnras/stt546}

\bibitem[{{Mummery} {et~al.}(2024){Mummery}, {Ingram}, {Davis}, \&
  {Fabian}}]{Mummery24}
{Mummery}, A., {Ingram}, A., {Davis}, S., \& {Fabian}, A. 2024, \mnras, 531,
  366, \dodoi{10.1093/mnras/stae1160}

\bibitem[{{Novikov} \& {Thorne}(1973)}]{NT73}
{Novikov}, I.~D., \& {Thorne}, K.~S. 1973, in Black Holes (Les Astres Occlus),
  ed. C.~{Dewitt} \& B.~S. {Dewitt} (Gordon and Breach: New York, NY), 343--450

\bibitem[{{Parker} {et~al.}(2016){Parker}, {Tomsick}, {Kennea}, {Miller},
  {Harrison}, {Barret}, {Boggs}, {Christensen}, {Craig}, {Fabian}, {F{\"u}rst},
  {Grinberg}, {Hailey}, {Romano}, {Stern}, {Walton}, \& {Zhang}}]{Parker16}
{Parker}, M.~L., {Tomsick}, J.~A., {Kennea}, J.~A., {et~al.} 2016, \apjl, 821,
  L6, \dodoi{10.3847/2041-8205/821/1/L6}

\bibitem[{{Petrucci} {et~al.}(2020){Petrucci}, {Gronkiewicz},
  {R{\'o}{\.z}a{\'n}ska}, {Belmont}, {Bianchi}, {Czerny}, {Matt}, {Malzac},
  {Middei}, {De Rosa}, {Ursini}, \& {Cappi}}]{Petrucci20}
{Petrucci}, P.~O., {Gronkiewicz}, D., {R{\'o}{\.z}a{\'n}ska}, A., {et~al.}
  2020, \aap, 634, A85, \dodoi{10.1051/0004-6361/201937011}

\bibitem[{{Plant} {et~al.}(2014){Plant}, {Fender}, {Ponti}, {Mu{\~n}oz-Darias},
  \& {Coriat}}]{Plant14}
{Plant}, D.~S., {Fender}, R.~P., {Ponti}, G., {Mu{\~n}oz-Darias}, T., \&
  {Coriat}, M. 2014, \mnras, 442, 1767, \dodoi{10.1093/mnras/stu867}

\bibitem[{{Poutanen} {et~al.}(2007){Poutanen}, {Lipunova}, {Fabrika},
  {Butkevich}, \& {Abolmasov}}]{Poutanen07}
{Poutanen}, J., {Lipunova}, G., {Fabrika}, S., {Butkevich}, A.~G., \&
  {Abolmasov}, P. 2007, \mnras, 377, 1187,
  \dodoi{10.1111/j.1365-2966.2007.11668.x}

\bibitem[{{Poutanen} \& {Svensson}(1996)}]{PS96}
{Poutanen}, J., \& {Svensson}, R. 1996, \apj, 470, 249, \dodoi{10.1086/177865}

\bibitem[{{Reis} {et~al.}(2008){Reis}, {Fabian}, {Ross}, {Miniutti}, {Miller},
  \& {Reynolds}}]{Reis08}
{Reis}, R.~C., {Fabian}, A.~C., {Ross}, R.~R., {et~al.} 2008, \mnras, 387,
  1489, \dodoi{10.1111/j.1365-2966.2008.13358.x}

\bibitem[{{Reis} {et~al.}(2013){Reis}, {Reynolds}, {Miller}, {Walton},
  {Maitra}, {King}, \& {Degenaar}}]{Reis13}
{Reis}, R.~C., {Reynolds}, M.~T., {Miller}, J.~M., {et~al.} 2013, \apj, 778,
  155, \dodoi{10.1088/0004-637X/778/2/155}

\bibitem[{{Rout} {et~al.}(2020){Rout}, {Vadawale}, \& {M{\'e}ndez}}]{Rout20}
{Rout}, S.~K., {Vadawale}, S., \& {M{\'e}ndez}, M. 2020, \apjl, 888, L30,
  \dodoi{10.3847/2041-8213/ab629e}

\bibitem[{{Salvesen} \& {Miller}(2021)}]{Salvesen21}
{Salvesen}, G., \& {Miller}, J.~M. 2021, \mnras, 500, 3640,
  \dodoi{10.1093/mnras/staa3325}

\bibitem[{{Schnittman} \& {Krolik}(2009)}]{Schnittman09}
{Schnittman}, J.~D., \& {Krolik}, J.~H. 2009, \apj, 701, 1175,
  \dodoi{10.1088/0004-637X/701/2/1175}

\bibitem[{{Shakura} \& {Sunyaev}(1973)}]{SS73}
{Shakura}, N.~I., \& {Sunyaev}, R.~A. 1973, \aap, 24, 337

\bibitem[{{Shakura} \& {Sunyaev}(1976)}]{Shakura76}
---. 1976, \mnras, 175, 613, \dodoi{10.1093/mnras/175.3.613}

\bibitem[{{S{\k{a}}dowski}(2009)}]{Sadowski09}
{S{\k{a}}dowski}, A. 2009, \apjs, 183, 171, \dodoi{10.1088/0067-0049/183/2/171}

\bibitem[{{S{\k{a}}dowski}(2011)}]{Sadowski11a}
---. 2011, PhD thesis, arXiv:1108.0396, \dodoi{10.48550/arXiv.1108.0396}

\bibitem[{{S{\k{a}}dowski} {et~al.}(2011){S{\k{a}}dowski}, {Abramowicz},
  {Bursa}, {Klu{\'z}niak}, {Lasota}, \& {R{\'o}{\.z}a{\'n}ska}}]{Sadowski11b}
{S{\k{a}}dowski}, A., {Abramowicz}, M., {Bursa}, M., {et~al.} 2011, \aap, 527,
  A17, \dodoi{10.1051/0004-6361/201015256}

\bibitem[{{Spruit}(2002)}]{Spruit02}
{Spruit}, H.~C. 2002, \aap, 381, 923, \dodoi{10.1051/0004-6361:20011465}

\bibitem[{{Sridhar} {et~al.}(2020){Sridhar}, {Garc{\'\i}a}, {Steiner},
  {Connors}, {Grinberg}, \& {Harrison}}]{Sridhar20}
{Sridhar}, N., {Garc{\'\i}a}, J.~A., {Steiner}, J.~F., {et~al.} 2020, \apj,
  890, 53, \dodoi{10.3847/1538-4357/ab64f5}

\bibitem[{{Straub} {et~al.}(2011){Straub}, {Bursa}, {S{\k{a}}dowski},
  {Steiner}, {Abramowicz}, {Klu{\'z}niak}, {McClintock}, {Narayan}, \&
  {Remillard}}]{Straub11}
{Straub}, O., {Bursa}, M., {S{\k{a}}dowski}, A., {et~al.} 2011, \aap, 533, A67,
  \dodoi{10.1051/0004-6361/201117385}

\bibitem[{{Svensson} \& {Zdziarski}(1994)}]{SZ94}
{Svensson}, R., \& {Zdziarski}, A.~A. 1994, \apj, 436, 599,
  \dodoi{10.1086/174934}

\bibitem[{{Svoboda} {et~al.}(2024){Svoboda}, {Dov{\v{c}}iak}, {Steiner},
  {Kaaret}, {Podgorn{\'y}}, {Poutanen}, {Veledina}, {Muleri}, {Taverna},
  {Krawczynski}, {Brigitte}, {Datta}, {Bianchi}, {Mu{\~n}oz-Darias}, {Negro},
  {Rodriguez Cavero}, {Castro Segura}, {Bollemeijer}, {Garc{\'\i}a}, {Ingram},
  {Matt}, {Nathan}, {Weisskopf}, {Altamirano}, {Baldini}, {Capitanio}, {Egron},
  {Emami}, {Hu}, {Marra}, {Mastroserio}, {Petrucci}, {Ratheesh}, {Soffitta},
  {Tombesi}, {Yang}, \& {Zhang}}]{Svoboda24}
{Svoboda}, J., {Dov{\v{c}}iak}, M., {Steiner}, J.~F., {et~al.} 2024, \apjl,
  966, L35, \dodoi{10.3847/2041-8213/ad402e}

\bibitem[{{Tauris}(2022)}]{Tauris22}
{Tauris}, T.~M. 2022, \apj, 938, 66, \dodoi{10.3847/1538-4357/ac86c8}

\bibitem[{{Wilms} {et~al.}(2000){Wilms}, {Allen}, \& {McCray}}]{Wilms00}
{Wilms}, J., {Allen}, A., \& {McCray}, R. 2000, \apj, 542, 914,
  \dodoi{10.1086/317016}

\bibitem[{{Wu} {et~al.}(2016){Wu}, {Xie}, {Yuan}, \& {Gan}}]{Wu16}
{Wu}, M.-C., {Xie}, F.-G., {Yuan}, Y.-F., \& {Gan}, Z. 2016, \mnras, 459, 1543,
  \dodoi{10.1093/mnras/stw742}

\bibitem[{{Zdziarski} {et~al.}(2024{\natexlab{a}}){Zdziarski}, {Banerjee},
  {Chand}, {Dewangan}, {Misra}, {Szanecki}, \&
  {Nied{\'z}wiecki}}]{Zdziarski24a}
{Zdziarski}, A.~A., {Banerjee}, S., {Chand}, S., {et~al.} 2024{\natexlab{a}},
  \apj, 962, 101, \dodoi{10.3847/1538-4357/ad1b60}

\bibitem[{{Zdziarski} \& {De Marco}(2020)}]{ZDM20}
{Zdziarski}, A.~A., \& {De Marco}, B. 2020, \apjl, 896, L36,
  \dodoi{10.3847/2041-8213/ab9899}

\bibitem[{{Zdziarski} {et~al.}(2020){Zdziarski}, {Szanecki}, {Poutanen},
  {Gierli{\'n}ski}, \& {Biernacki}}]{Z20_thcomp}
{Zdziarski}, A.~A., {Szanecki}, M., {Poutanen}, J., {Gierli{\'n}ski}, M., \&
  {Biernacki}, P. 2020, \mnras, 492, 5234, \dodoi{10.1093/mnras/staa159}

\bibitem[{{Zdziarski} {et~al.}(2019){Zdziarski}, {Zi{\'o}{\l}kowski}, \&
  {Miko{\l}ajewska}}]{Zdziarski19a}
{Zdziarski}, A.~A., {Zi{\'o}{\l}kowski}, J., \& {Miko{\l}ajewska}, J. 2019,
  \mnras, 488, 1026, \dodoi{10.1093/mnras/stz1787}

\bibitem[{{Zdziarski} {et~al.}(2024{\natexlab{b}}){Zdziarski}, {Chand},
  {Banerjee}, {Szanecki}, {Janiuk}, {Lubi{\'n}ski}, {Nied{\'z}wiecki},
  {Dewangan}, \& {Misra}}]{Zdziarski24b}
{Zdziarski}, A.~A., {Chand}, S., {Banerjee}, S., {et~al.} 2024{\natexlab{b}},
  \apjl, 967, L9, \dodoi{10.3847/2041-8213/ad43ed}

\end{thebibliography}

\end{document}